\documentclass[preprint]{revtex4}
\usepackage{amsmath,amssymb,epic,eepic}
\usepackage[dvips]{graphicx}
\usepackage{epsfig}
\usepackage{color}

\begin{document}

\title{Subnanosecond single electron source in the time-domain
}


\author{A. Mah\'{e}$^{(1)}$ }
\author{F.D. Parmentier $^{(1)}$}
\author{G. F\`{e}ve  $^{(1)}$}
\author{J.-M. Berroir$^{(1)}$ }
\author{ T. Kontos $^{(1)}$}
\author{ A. Cavanna $^{(2)}$}
\author{B. Etienne $^{(2)}$}
\author{ Y. Jin$^{(2)}$}
\author{D.C. Glattli$^{(1,3)}$}
\author{ B. Pla\c{c}ais$^{(1)}$ }


\affiliation{ (1) Laboratoire Pierre Aigrain, D{\'e}partement de
Physique de l'Ecole Normale Sup\'erieure, 24 rue Lhomond, 75231
Paris Cedex 05, France}
  \affiliation{(2)  Laboratoire de Photonique et
Nanostructures, UPR20 CNRS, Route de Nozay, 91460 Marcoussis
Cedex, France }
 \affiliation{(3) Service de Physique de
l'Etat Condens{\'e}, CEA Saclay, F-91191 Gif-sur-Yvette, France }

\begin{abstract}
We describe here the realization of a single electron source
similar to single photon sources in optics. On-demand single
electron injection is obtained using a quantum dot connected to
the conductor via a tunnel barrier of variable transmission
(quantum point contact). Electron emission is triggered by a
sudden change of the dot potential which brings a single energy
level above the Fermi energy in the conductor. A single charge is
emitted on an average time ranging from 100 ps to 10 ns ultimately
determined by the barrier transparency and the dot charging
energy. The average single electron emission process is recorded
with a 0.5 ns time resolution using a real-time fast acquisition
card. Single electron signals are compared to simulation based on
scattering theory approach adapted for finite excitation energies.

\end{abstract}

\maketitle

\section{Introduction}

\label{intro} The controlled emission of single photons in quantum
optics has opened a new route for a quantum physics based on
entanglement of several photons \cite{Gisin_Rev02,Kok_Rev07}.
Recently, a similar on-demand injection of single electrons in a
quantum conductor at a well defined energy, whose uncertainty is
ultimately controlled by the tunneling rate has been realized
\cite{Feve2007,FeveEP2DS}. This experimental realization opens new
opportunity for quantum experiments with single electrons
\cite{collider2008}, including flying qubits in ballistic conductors
as envisaged in Ref.
\cite{flying_quBits1,flying_quBits2,flying_quBits3}. Of particular
relevance is the experiment described in Ref.\cite{collider2008}
which is an electron collider requiring accurate synchronization of
two coherent single electron sources.

We describe here experimental aspects on the realization of such a
single electron source \cite{Feve2007} and include extended results.
Single electron injection is triggered by a sudden variation of the
potential of a quantum dot. The electron sitting on the last
occupied energy level is then emitted in the lead in a coherent
wavepacket, with an energy width limited by the emission time which
can be tuned by the QPC transmission $D$ from 100 picoseconds to 10
nanoseconds.

The circuit, sketched in Fig.1 a), is realized in a 2D electron gas
made in a GaAsAl/GaAs heterojunction of nominal density $n_s = 1.7
\times 10^{15}\; \mathrm{m^{-2}}$ and mobility $\mu = 260\;
\mathrm{m^{2}V^{-1}s^{-1}}$. A quantum dot, of submicron dimensions,
is electrostatically coupled to a metallic top gate, located
$100\;\mathrm{nm}$ above the 2DEG, whose ac voltage amplitude,
$V_{exc}$, controls the dot potential at the subnanosecond
timescale. The dot is coupled to an electronic reservoir by a
quantum point contact (QPC) acting as a tunnel barrier of
transmission $D$ controlled by the gate voltage $V_g$. The series
addition of both elements constitutes what we call a mesoscopic
capacitor or equivalently a quantum RC circuit which is the most
basic AC-coupled device exhibiting non-trivial coherent transport
properties. For all measurements, the electronic temperature is
found about $200\;\mathrm{mK}$ for a base fridge temperature in the
range $30-70\;\mathrm{mK}$ and a magnetic field $B\approx 1.3
\;\mathrm{T}$ is applied to the sample so as to work in the quantum
Hall regime (filling factor $\nu=4$) with no spin degeneracy. We
think that the difference between electronic and fridge temperatures
mainly comes from unperfect gate voltage filtering. $V_g$ is tuned
to control the transmission $D$ of a single edge state from the
reservoir to the dot. It also controls by capacitive coupling the
mean potential of the dot.

\begin{figure}
\includegraphics[scale=0.4]{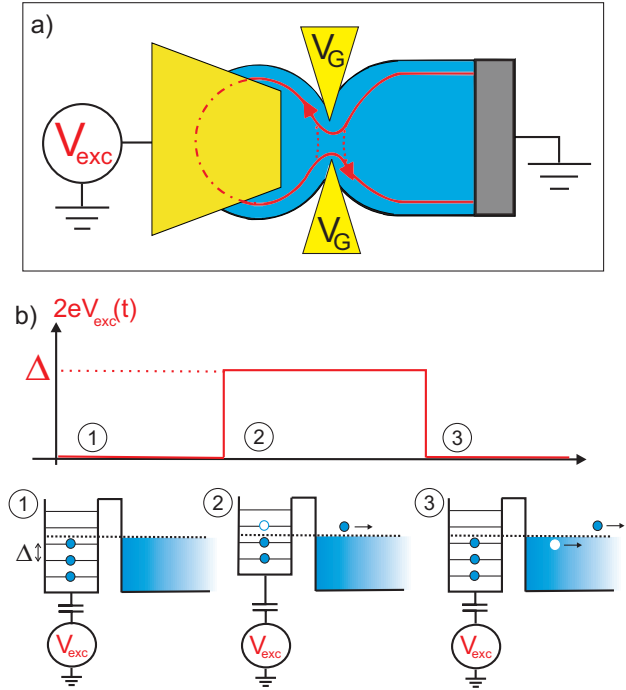}
\caption{ a) Schematic of the circuit. b) Schematic of single
charge injection.}
\end{figure}

In the linear response regime to a high frequency potential
excitation applied to the top gate, the mesoscopic circuit forms a
quantum RC circuit. Its study of the charge relaxation is described
in \cite{Gabelli06Science,Gabelli2005} where it was demonstrated the
quantization of the charge relaxation resistance $R_q=h/2e^2$ for a
single mode conductor with no spin degeneracy predicted ten years
before \cite{BPT93PL,BPT93PRL,PTB96PRB}. The total capacitance of
the dot is the series addition of the geometrical capacitance $C$
and the quantum capacitance $C_q=e^2N(\epsilon_f)$ where
$N(\epsilon_f)$ is the density of states of the dot at the Fermi
energy. As in our experiment we have $C\gg C_q$, the charge
relaxation time $\tau_q\simeq R_qC_q = hN(\epsilon_f)/2$ is a new
indirect measurement of the density of states. Importantly for the
present work, the linear response allows for a precise extraction of
the parameters of the dot : level spacing
$\Delta\approx2\;\mathrm{K}$, geometrical capacitance $C \approx
3\;\mathrm{fF}$, and the total addition energy
$E_c=\Delta+e^2/C=2.5\;\mathrm{K}$. As capacitance effects are
dominated by the quantum capacitance $C_q$, we shall neglect the
geometrical capacitance in the rest of the paper and take
$E_c\equiv\Delta$. Kinetic inductance effects in the leads which
give rise to additional time delays \cite{Gabelli2007,Wang2007}
which can be neglected in our small structures. Also we disregard
here possible decoherence effects beside thermal smearing due to
finite reservoir temperature. For a theoretical discussion on
interactions and finite
coherence time effects on the charge relaxation process the reader is referred to Refs.\cite{Nigg2006,Nigg2008}.\\

This paper deals with the regime of high excitation amplitudes
comparable with the level spacing $\Delta\approx 200\;\mathrm{\mu
eV}$. By applying a sudden step voltage on the top gate, the first
occupied energy level of the dot is brought above the Fermi energy
of the reservoir and a single electron is emitted (see Fig. 1 b)) on
a characteristic time $\tau$ which is expected to be related to the
width of this single energy level, $\tau = h/D\Delta$. In this
experiment the emission time, $\tau = h/D\Delta$, was larger than
the pulse rise time ($\sim50 \;\mathrm{ps}$) so that the lifting of
the energy level can be regarded as instantaneous.  This emission
time $\tau$ differs in general from the above charge relaxation time
$\tau_q$ of the coherent regime. However, it coincides with the
charge relaxation time in the incoherent regime as observed at high
temperatures and/or low transmissions ($k_B T>> D \Delta$)
\cite{Gabelli06Science}. Multiple charge emission is prevented by
both Coulomb interactions and Pauli exclusion principle. In this
work single charge detection is achieved by statistical averaging of
a large number of events. To repeat the experiment, the dot needs to
be reloaded by putting the potential back to its initial value. One
electron is then absorbed by the dot, or equivalently a hole is
emitted in the Fermi sea. Periodic repetition of square voltage
excitations then generates periodic emission of single electron-hole
pairs which leads to a quantized ac-current in units of $2ef$
\cite{Feve2007,FeveEP2DS}.
This is a marked difference with pumps which show quantization of the dc current in units of $ef$ \cite{Giblin2007}. \\

This periodic current can be measured either by phase resolved
harmonic measurements \cite{Feve2007} or directly in the time domain
with a fast acquisition and averaging card (Acqiris AP240 2GSa/s).
In this paper, we will focus on this second measurement scheme.

\section{Current pulses in the time domain}
\label{sec:1}

The average current generated by single charge transfer is detected
in time domain by the voltage drop on a $50 \;\mathrm{\Omega}$
resistor located at the input of a broadband low noise cryogenic
amplifier. Given an escape time $ \tau \approx 500\;\mathrm{ps}$,
the input voltage amplitude is $V \approx (50 \Omega) \times e/\tau
\approx 16 \;\mathrm{nV}$. For a $15\;\mathrm{K}$ noise temperature
amplifier in a $1GHz$ bandwidth, the input noise amplitude is a few
$\mathrm{\mu V}$. Single shot measurement of single charges is thus
out of reach. Only statistical measurements of the average current
can be achieved. The experiment needs to be repeated about
$10^5$ times to restore a signal to noise ratio close to unity with our current setup. \\

Although each electron is detected at a well defined time, single
electron emission is a quantum probabilistic process. We thus expect
that the average current resulting from the accumulation of a large
set of single electron events will reconstruct the probability
density of electronic emission. As in a usual decay process, it
should follow an exponential relaxation on a characteristic time
given by the escape time $\tau$. This exponential relaxation can be
viewed in the lumped elements language as the mere relaxation time
of a RC circuit. A current pulse with opposite sign is expected for the
single hole emission.\\

The expected current can be theoretically calculated by a scattering
theory approach similar to that of \cite{PTB96PRB} extended to high
excitations amplitudes, while neglecting interactions as discussed
above \cite{Feve2007}. Further developments include estimation of
source quantization accuracy and noise \cite{Moskalets2008} and the
emission of secondary electron-hole pairs
\cite{Keeling2008,Levitov2006}. As a single edge state is
transmitted to the dot, we will consider below a single mode
conductor with no spin degeneracy. Odd harmonics of the current
$I_{(2k+1)\omega}$ can be written as \cite{FeveThesis2006}

\begin{eqnarray}\label{i2k+1}
I_{(2k+1)\omega} & =  \frac{e}{2 \pi i h } \frac{1}{2k+1}\int &
d\epsilon
 [1-s^{+}(\epsilon)s(\epsilon+(2k+1)\hbar \omega) ]\times  \nonumber \\
& &
[f(\epsilon+(2k+1)\hbar\omega)+f(\epsilon)-f(\epsilon-2eV_{exc}+(2k+1)\hbar
\omega)-f(\epsilon-2eV_{exc})] \qquad ,
\end{eqnarray}
where $2V_{exc}$ is the peak to peak amplitude of the excitation
voltage and $s(\epsilon)$ is the energy-dependent scattering matrix.
At low frequency $(2k+1)\omega << 1/\tau$, the current can be
expanded up to the second order in $(2k+1)\omega$:
\begin{eqnarray}
I_{(2k+1)\omega} = \frac{i2V_{exc}}{\pi (2k+1)} \int d\epsilon
\Big[-i(2k+1)\omega e^{2}N(\epsilon) +
\frac{h}{2e^{2}}[e^{2}N(\epsilon)(2k+1)\omega]^{2} \Big] \;
\frac{f(\epsilon-2eV_{exc})-f(\epsilon)}{2eV_{exc}} \quad \;
\label{current_approx}\end{eqnarray} where the prefactor is the
$(2k+1)^{th}$ harmonic of the excitation voltage, and the density of
states $N(\epsilon)$ is related to the scattering matrix through
$N(\epsilon)=1/(2i\pi)\times s(\epsilon)ds(\epsilon)/d\epsilon$. The
above equation thus establishes that the circuit is equivalent to an
RC circuit with a $V_{exc}$-dependent capacitance and resistance
given by :
\begin{eqnarray}
\widetilde{C_{q}}&=& e^{2}\int d\epsilon
N(\epsilon)\frac{f(\epsilon-2eV_{exc})-f(\epsilon)}{2eV_{exc}} \label{Cqnl}\\
\widetilde{R_{q}} & =& \frac{h}{2e^{2}} \frac{ \int d\epsilon
N(\epsilon)^{2}\frac{f(\epsilon-2eV_{exc})-f(\epsilon)}{2eV_{exc}}}{[\int
d\epsilon
N(\epsilon)\frac{f(\epsilon-2eV_{exc})-f(\epsilon)}{2eV_{exc}}]^{2}}
\label{Rqnl}.
\end{eqnarray}
In the time domain, this corresponds to an exponentially decaying
current :
\begin{eqnarray}
I(t) = \frac{q}{\tau} e^{-t/\tau}, \;\rm{with}\;& q =
\widetilde{C_{q}}\times 2V_{exc}, \;\rm{and}\; &
 \tau = \widetilde{R_{q}}\widetilde{C_{q}}
\label{expo}\end{eqnarray}

Equation (\ref{Cqnl}) shows that the charge $q$ is given by the
density of states integrated between the extremum values of the
square excitation. If one energy level initially located below the
Fermi energy is put above, then a single charge is transferred as
naively predicted. In the limit of low transmissions $D<<1$,
Eqs.(\ref{Cqnl}) and (\ref{Rqnl}) give a relaxation time $\tau
\approx
h/D\Delta$ as anticipated.\\

\begin{figure}
\includegraphics[scale=0.5]{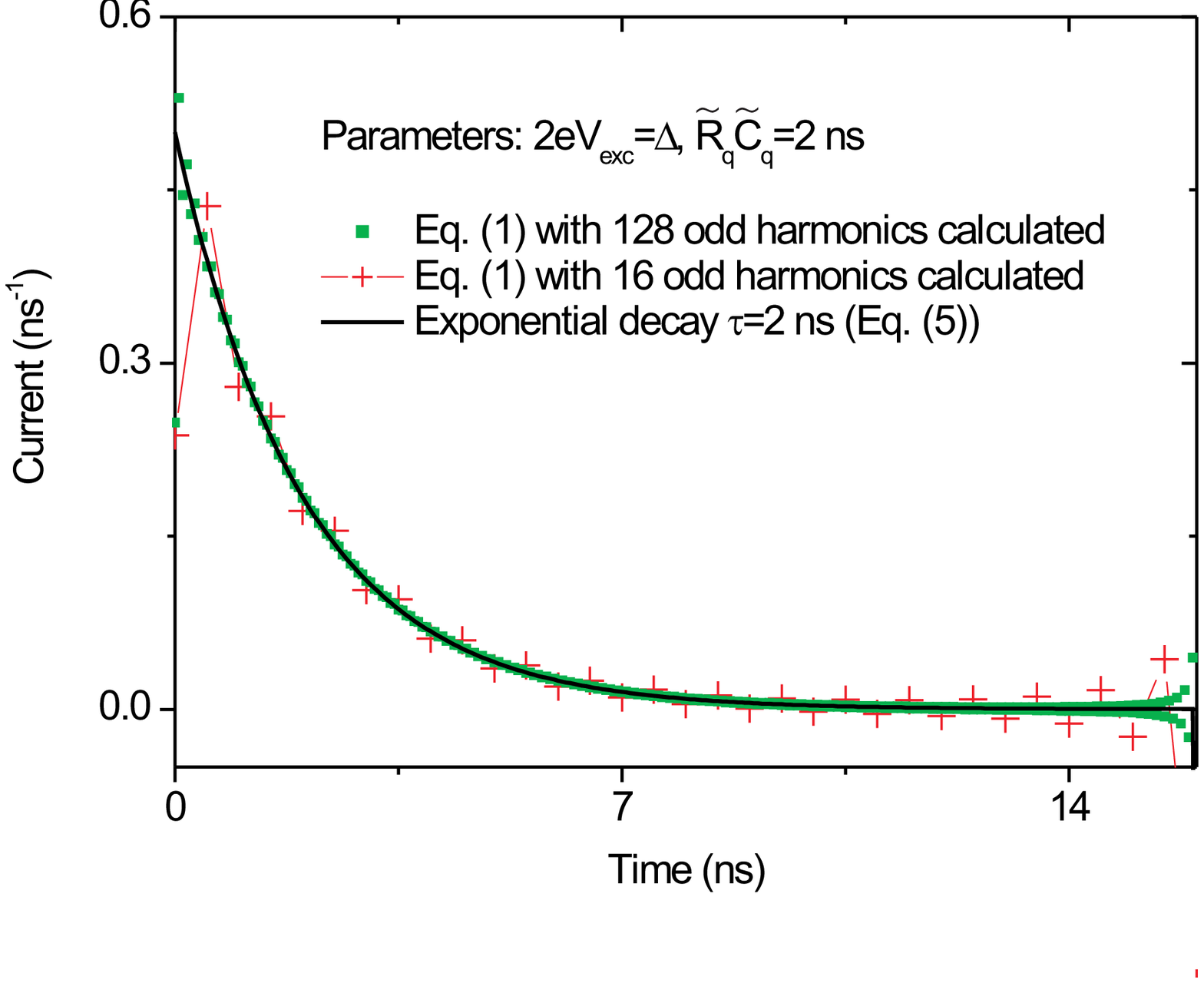}
\caption{  Comparison between Eq.(\ref{i2k+1}) and Eq.(\ref{expo})
for the calculation of the current. Both expressions coincide when
a large number of harmonics are calculated. Deviations at short
times are observed when only 16 harmonics are taken into account.
} \label{figure2}\end{figure}

Comparison in Figure \ref{figure2}  between the exact expression of
the current Eq.(\ref{i2k+1}) and its  low frequency limit,
Eq.(\ref{current_approx}) and (\ref{expo}) (escape time $\tau
=2\;\mathrm{ns}$), validates the low-frequency approximation.
Therefore, we shall only consider
this simpler form from now on.\\

Even harmonics can also be calculated although their expressions are
more involved. They reveal the possible differences between the
electron and hole emission processes which are suppressed in our
measurement procedure as discussed below. \\

Experimentally one can only access a finite number of  current
harmonics. In our case, the current is recorded using a $1 GHz$ fast
averaging card. With a drive frequency of $31.25\;\mathrm{MHz}$,
$32$ harmonics can be measured (16 odd harmonics). One can see in
Fig. \ref{figure2} that this bandwidth limitation hardly affects the
 relaxation process for $\tau\geq 1\;\mathrm{ns}$. \\

The measurement setup will now be detailed in the next section.

\section{Experimental Setup}

\begin{figure}
\includegraphics[scale=0.8]{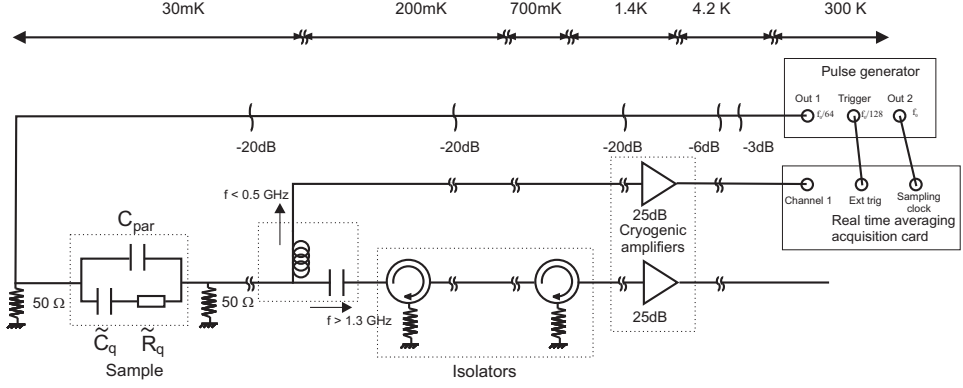}
\caption{Schematic of the experimental setup.}
\label{figure3}\end{figure}

An input square excitation of approximately $1\;\mathrm{V}$
amplitude and $50 \;\mathrm{ps}$ risetime is provided by a pulse
generator at frequency $31.25 \;\mathrm{MHz}$ and fed at the input
of a $40 \;\mathrm{GHz}$ rf-transmission line. After an attenuation
of $\approx 70\;\mathrm{dB}$, it reduces to a square signal of a few
hundreds of microvolts applied to a $50 \;\mathrm{\Omega}$ resistor
at the gate of the sample, see Fig.\ref{figure3}. Two ultra low
noise cryogenic amplifiers record the voltage drop on another $50
\;\mathrm{\Omega}$ resistor located at the output of the sample. A
bias tee separates the low frequency ($< 500\;\mathrm{MHz}$) part of
the current signal from the high frequency ($>1.3 \;\mathrm{GHz}$)
part. The high frequency line is used to measure the harmonic
response of the sample for a high frequency drive (typically $1.5
\;\mathrm{GHz}$) as described in Ref.\cite{Feve2007}. We will focus
here on the low frequency part from which we extract the time domain
dependence of the current for a $31.25 \;\mathrm{MHz}$ drive. The
first measurement scheme gives access to ultra short times ($\approx
10 \;\mathrm{ps}$ resolution) whereas the second method gives the
complete time dependence of the current but on longer times ($> 500
\;\mathrm{ps}$). Note that all the spectrum cannot be measured by a
single amplifier as the sample is protected from the high frequency
part of the amplifiers current noise by the use of rf-isolators with
limited bandwidth
$1.2-1.8 \;\mathrm{GHz}$. \\

A fast averaging card (Acqiris AP240) of $1\;\mathrm{GHz}$ bandwidth
(sampling time $500 \;\mathrm{ps}$) records the output current.
Averaging over long times (a few seconds) requires a perfect
synchronisation of the sampling clock with the drive frequency.
Therefore, the sampling clock is generated by the generator and the
drive is an integer fraction of the clock:
$f_{drive}=2\;\mathrm{GHz}/64=31.25 \;\mathrm{MHz}$ (see
Fig.\ref{figure3}). One run of measures is triggered every $4096$
periods of the drive (so that one run lasts $100 \;\mathrm{\mu s}$).
Typically $65 000 $ runs are then averaged by the card in real time
for a total measurement time of $\approx 10 \;\mathrm{s}$. The
overall procedure gives a signal to
noise ratio $\sim 10$ in a $1 \;\mathrm{GHz}$ bandwidth.  \\

As mentioned before, the effective bandwidth of the detection line
is limited to $500 \;\mathrm{MHz}$ by the first bias tee. The
cryogenic amplifiers also suffer from a low frequency cutoff of a
few tens of MHz. Moreover, impedance mismatch of the amplifiers
leads to multiple reflections (echoes) which also affect the
measured time dependence of the signal. Finally the current signal
is affected by bandwidth limitations and distortion.  However, these
spurious effects can by corrected by proper calibration.\\

\begin{figure}
\includegraphics[scale=0.6]{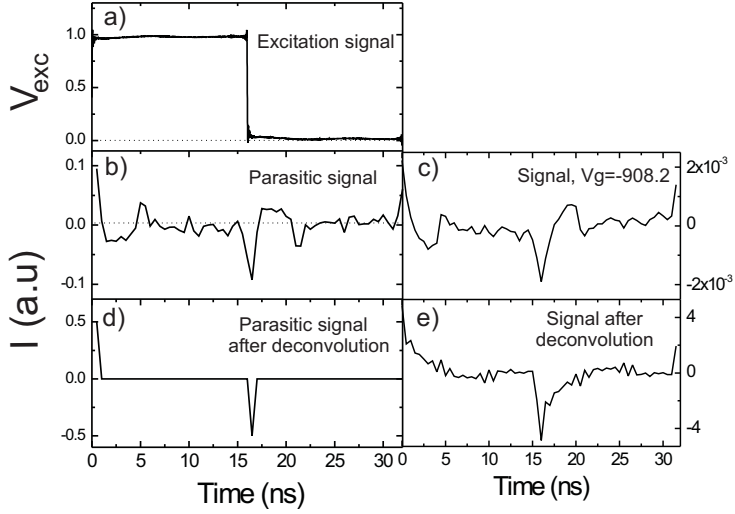}
\caption{Excitation, parasitic and  sample contribution signals (a),
 (b)and (c). Panels (d) and (e) show the parasitic and sample signals
 after deconvolution of the amplification line response.} \label{figure4}\end{figure}

As seen previously, the sample can be represented by the series
addition of a resistance and capacitance given by Eqs.(\ref{Cqnl})
and (\ref{Rqnl}). However, it is always bypassed by a parasitic
capacitive coupling $C_{par}$, see Fig.\ref{figure3}. In our setup
it corresponds to the largest part of the transmitted current
($C_{par}\approx 50\;\mathrm{fF}$ compared to $\widetilde{C}_q
\approx 0.7\;\mathrm{fF}$). By varying the gate voltage $V_g$, the
quantum dot can be decoupled from the electronic reservoir (at
transmission $D=0$, $\widetilde{R}_q= \infty $ ), allowing for an
accurate measurement of the parasitic contribution which is then
subtracted to only keep the sample contribution. This parasitic
signal can be used as a reference signal for the detection line.
Indeed for a capacitive coupling, the voltage drop on the output $50
\;\mathrm{\Omega}$ resistor is given by an exponential relaxation on
time $RC << 500\;\mathrm{ps}$. It can thus be approximated by a
Dirac delta function compared to the $500\;\mathrm{ps}$ sampling
time of the card. Therefore, the fourier transform of the parasitic
contribution gives an accurate measurement of the odd fourier
components of the detection line bandwidth.  The pristine signal can
be reconstructed by dividing its odd measured fourier components by
those of the detection line. As even components cannot be
calibrated, they are disregarded in this experiment. This amounts to
disregard differences in the electron and hole emission processes. A
typical measured parasitic signal is given on Fig.\ref{figure4}.b
where one can see the previously mentioned distortions (widening
caused by the $500 \;\mathrm{MHz}$ bandwidth, drop to negative
values of current after a positive peak caused by the low frequency
cutoff and echoes). The effect of the deconvolution process on a
typical trace can then be seen on Fig.\ref{figure4}.e, where most of
the distortions have been erased and one recovers an
exponential decrease of the current. \\

The next section will now describe the main results obtained with
this measurement setup.

\section{Results}

\begin{figure}
\includegraphics[scale=0.6]{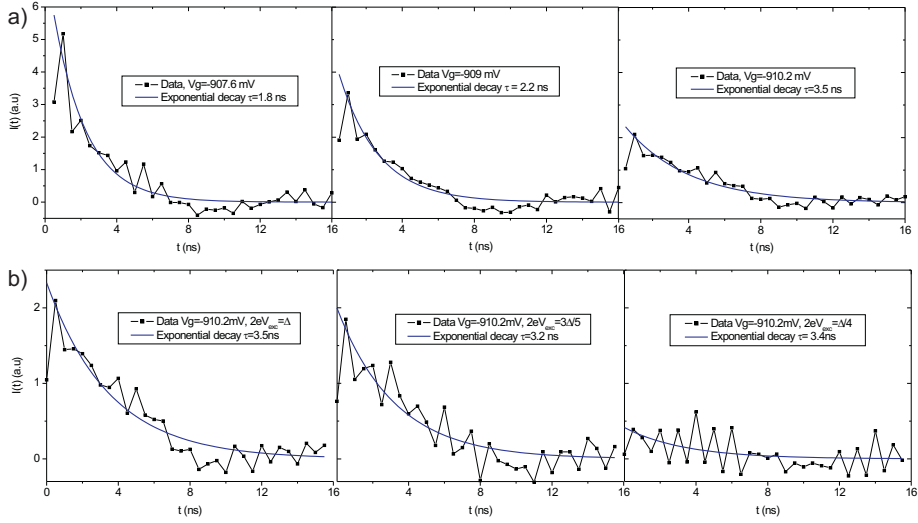}
\caption{ a) Data and exponential fits for $V_g = -907.6, -909$
and $-910.2\;\mathrm{mV}$ and an excitation amplitude of
$2eV_{exc}=\Delta$. b) Data and exponential fits at
$V_g=-910.2\;\mathrm{mV}$ for $2eV_{exc}= \Delta$, $3\Delta/5$ and
$\Delta/4$. } \label{figure5}\end{figure}

Fig. \ref{figure5}.a presents a few typical current pulses obtained
for $2eV_{exc} \approx \Delta$ for different values of the gate
voltage corresponding to different transmissions. For lower values
of the gate voltage, the transmission decreases and the escape time
increases as predicted by Eq.(\ref{expo}). Moreover, all these
curves are well fitted by an exponential relaxation from which the
escape time $\tau$ can be extracted. $\tau$ can be continuously
varied within two orders of magnitude, from a hundred of picoseconds
to ten nanoseconds by a simple shift in the gate voltage $V_g$. By
contrast, in this regime of low transmissions, the escape time does
not depend much on the excitation amplitude as can be seen on
Fig.\ref{figure5}.b.  Again this behavior is expected as for $h/\tau
<<k_B T$, the linear regime relaxation time $\tau_q$ averaged on the
energy window $k_BT$ coincides with the high excitation value
$h/D\Delta$.

\begin{figure}
\includegraphics[scale=0.5]{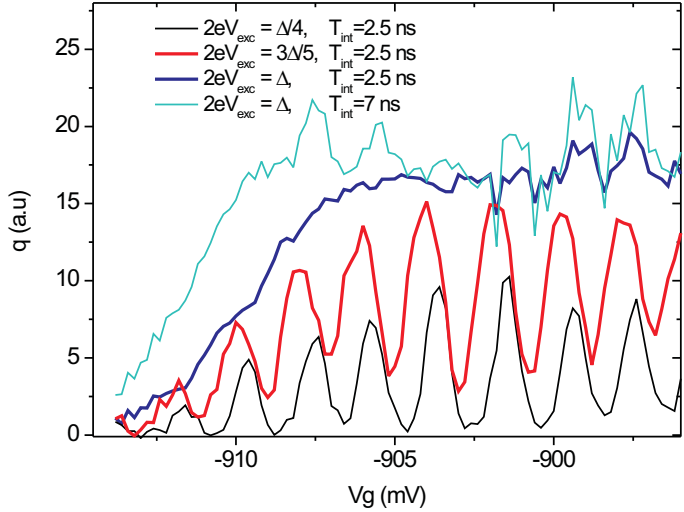}
\caption{ Average charge transferred in the first
$2.5\;\mathrm{ns}$ for $2eV_{exc}=\Delta/4$, $3\Delta/5$ and
$\Delta$ and after $T_{int}=7\;\mathrm{ns}$ for
$2eV_{exc}=\Delta$.} \label{figure6}\end{figure}

The averaged transmitted charge in a given time can be obtained by
integrating the current pulse over a given time window.
Fig.\ref{figure6} represents the average transmitted charge in a
window of $2.5\;\mathrm{ns}$ for three excitation amplitudes
$2eV_{exc} = \Delta/4, 3\Delta/5, \Delta$. For $2eV_{exc} =
\Delta/4$, the transmitted charge exhibits strong oscillations with
gate voltage. In some cases, the excitation amplitude is not high
enough to promote an energy level above the Fermi energy and the
transferred charge is zero. On the opposite, when a level is close
to resonance the transmitted charge shows a peak. When the
excitation amplitude is increased these structures tend to disappear
up to the level $2eV_{exc} = \Delta$ for which the transmitted
charge does not depend on gate voltage anymore (except for low
values of gate voltage as the escape time becomes longer than the
integration time). In this case one electron is transferred
independently of the initial value of the dot potential. It is the
time-domain counterpart of the quantized ac current $I_\omega=2ef$
in the harmonic measurement reported in Ref.\cite{Feve2007}. One can
also see that the peaks observed for $2eV_{exc}= 3 \Delta/5$ are
very close to the curve obtained for $2eV_{exc}=\Delta$. In this
case, the transmitted charge has a very small dependence on the
excitation amplitude. It exhibits a quantized plateau as only a
single charge can be emitted at each period of the drive. One can
also see that for a longer integration window of $7\;\mathrm{ns}$,
the curve obtained for $2eV_{exc}=\Delta$ is shifted to lower
transmissions as expected. At the same time, high transmission part
of the curve becomes more noisy as the integration window
incorporates more noise.

\begin{figure}
\includegraphics[scale=0.5]{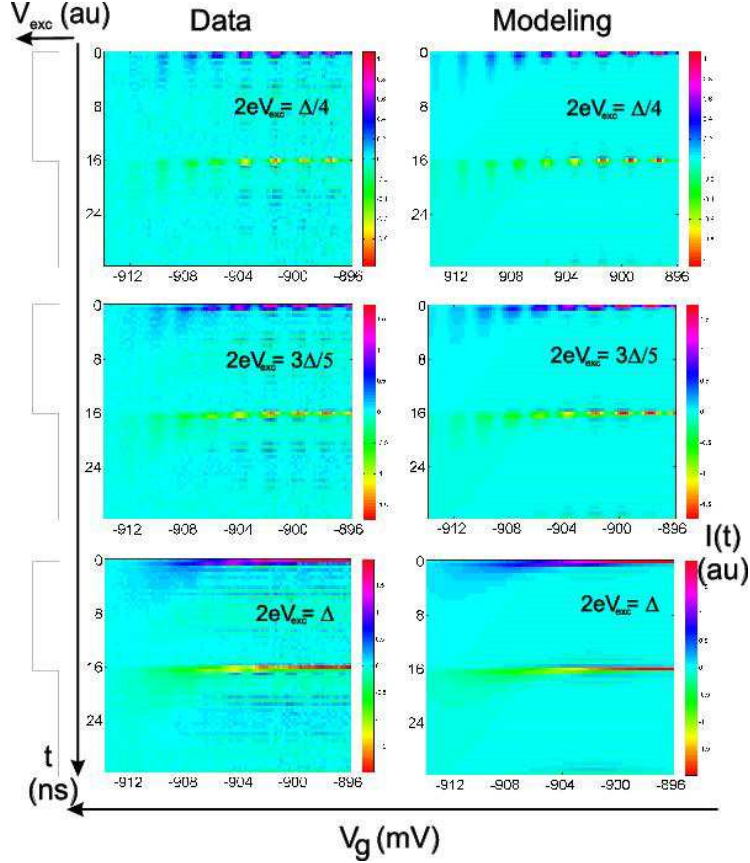}
\caption{Two-dimensional plot of the current represented in
colorscale as a function of time (vertical axis) and gate voltage
(horizontal axis). The current is plotted for $2eV_{exc}= \Delta/4$,
$3\Delta/5$ and $\Delta$ experimental data are plotted and the left
side, numerical simulations on the right. Excitation pulses are
displayed on the left so as to emphasize synchronization of
electron/hole emission with the rising/falling edge of the pulses.
Difference between experiment and model lies mainly in the presence
of ripples at finite time delays ion the former which we attribute
to imperfections in the deconvolution process. }
\label{figure7}\end{figure}

Finally all the results obtained for different values of the gate
voltage and excitation amplitude can be represented on a
two-dimensional colorplot, Fig.\ref{figure7},  where the current is
represented in colorscale. We observe again at low amplitudes the
periodic oscillations reflecting the periodicity of the dot density
of states (horizontal axis).  Broadening of the current peaks occurs
at low transmission as the escape time becomes longer (vertical
axis). For $2eV_{exc}=\Delta$, the oscillations vanish as one
electron is emitted at each period and one can only observe the
variation of escape time with gate voltage. The right part of
Fig.\ref{figure7} represents numerical simulations using
Eq.(\ref{i2k+1}) with no adjustable parameters (level spacing and
transmission are extracted from the linear measurements at low
amplitudes). The experimental behavior is very well reproduced in
the numerical simulations in the nanosecond  scale of the time
resolved experiment.

\section{Conclusion}

We have reported here on the measurement in the time domain of time
controlled single electron emission using fast acquisition and
averaging techniques. Such a single electron source together with
single electron detection \cite{Feve2008} would open the way to
quantum electron optics experiments probing electron antibunching or
electron entanglement \cite{collider2008,Hassler2007}.

\section{Acknowledgment}

The Laboratoire Pierre Aigrain is the CNRS-ENS mixed research unit
(UMR8551) associated with universities Paris 6 and Paris 7. The
research has been supported by the ANR-05-NANO-028 contract.





\end{document}